\documentclass[iop]{emulateapj}
\usepackage{apjfonts}
\usepackage{amssymb,latexsym,amsmath,graphics}
\slugcomment{} \shorttitle{A new correlation with lower kHz QPO}
\shortauthors{Erkut et al.}

\begin{document}

\title{A new correlation with lower kilohertz quasi-periodic oscillation frequency in the ensemble of low-mass X-ray binaries}

\author{M. Hakan Erkut,\altaffilmark{1,2} \c{S}ivan Duran,\altaffilmark{3} \"{O}nder \c{C}atmabacak,\altaffilmark{4} and Onur \c{C}atmabacak\altaffilmark{4}}

\altaffiltext{1}{Department of Physics, Faculty of Science and
Letters, Istanbul K\"{u}lt\"{u}r University,
Bak\i rk\"{o}y 34156, Istanbul, Turkey; m.erkut@iku.edu.tr}

\altaffiltext{2}{Feza G\"{u}rsey Center for Physics and
Mathematics, Bo\u{g}azi\c{c}i University, 34684, \c{C}engelk\"{o}y,
Istanbul, Turkey}

\altaffiltext{3}{Department of Astronomy and Space Sciences,
Graduate School of Science and Engineering, Istanbul University,
34116, Beyaz\i t, Istanbul, Turkey}

\altaffiltext{4}{Faculty of Engineering and Natural Sciences,
Sabanc\i\ University, 34956, Orhanl\i, Tuzla, Istanbul, Turkey}

\begin{abstract}
We study the dependence of kHz quasi-periodic oscillation (QPO) frequency on accretion-related parameters in the ensemble of neutron star low-mass X-ray binaries. Based on the mass accretion rate, $\dot{M}$, and the magnetic field strength, $B$, on the surface of the neutron star, we find a correlation between the lower kHz QPO frequency and $\dot{M}/B^{2}$. The correlation holds in the current ensemble of Z and atoll sources and therefore can explain the lack of correlation between the kHz QPO frequency and X-ray luminosity in the same ensemble. The average run of lower kHz QPO frequencies throughout the correlation can be described by a power-law fit to source data. The simple power-law, however, cannot describe the frequency distribution in an individual source. The model function fit to frequency data, on the other hand, can account for the observed distribution of lower kHz QPO frequencies in the case of individual sources as well as the ensemble of sources. The model function depends on the basic length scales such as the magnetospheric radius and the radial width of the boundary region, both of which are expected to vary with $\dot{M}$ to determine the QPO frequencies. In addition to modifying the length scales and hence the QPO frequencies, the variation in $\dot{M}$, being sufficiently large, may also lead to distinct accretion regimes, which would be characterized by Z and atoll phases.
\end{abstract}


\section{Introduction}\label{intr}

The kilohertz quasi-periodic oscillations (kHz QPOs) usually appear as two
simultaneous peaks in the $200-1300$ Hz range in the power spectra
of low-mass X-ray binaries (LMXBs) harboring neutron stars with spin
(or burst) frequencies in the $\sim 200-600$ Hz range
\citep{Klis2000}. The separation between two kHz QPO peaks is
roughly around $200-350$ Hz for almost all sources of different
spectral type and X-ray luminosity \citep{MB2007}.

There are two main spectral types of LMXB sources, the so-called Z
and atoll sources, which can be identified from the particular
shapes of their tracks in X-ray color-color and hardness-intensity
diagrams \citep{HK1989}. Among them, Z sources seem to be the
brightest sources in X-rays with the X-ray luminosity,
$L_{\mathrm{X}}$, close to the Eddington limit, $L_{\mathrm{E}}$,
whereas atoll sources are in the $0.005-0.2$ $L_{\mathrm{E}}$ range
\citep{Ford2000}.

The frequency range of kHz QPOs has been observed to be similar in sources
with different X-ray luminosities. The distribution of sources in the form
of parallel like groups or \emph{parallel tracks} can be seen in the QPO
frequency versus X-ray luminosity plot where sources with $L_{\mathrm{X}}$
close to the Eddington luminosity $L_{\mathrm{E}}$ may cover the same
frequency range, e.g., $500-1000$ Hz range, as compared to sources with $L_{%
\mathrm{X}}\approx 10^{-2}L_{\mathrm{E}}$ \citep{Ford2000}.

The \emph{parallel tracks} phenomenon has also been observed in a plot of kHz
QPO frequency versus X-ray count rate for individual sources
\citep{Zhang1998,Mendez1999,MK1999,Mendez2000,Klis2000}. The
correlation between X-ray flux and kHz QPO frequencies can be
clearly seen on short time scales such as hours or less than a day.
On longer time scales (more than a day), however, kHz QPO sources
are observed to follow different correlations which appear as
\emph{parallel tracks} in the plane of kHz QPO frequency versus
X-ray flux.

Effects such as anisotropic emission, source inclination, outflows
and two-component flow in a given source may play role in the
decoupling between $L_{\mathrm{X}}$ and the mass accretion rate,
$\dot{M}$, and therefore between the kHz QPO frequencies, $\nu _{\mathrm{kHz}}$,
and $L_{\mathrm{X}}$ if $\nu _{\mathrm{kHz}}$ is only tuned by $\dot{M}$
\citep{Wijnands1996,Ford2000,Mendez2001}. An explanation for the \emph{parallel tracks} phenomenon was proposed by \citet{Klis2001}. According to the scenario, the QPO frequency is set by the mass inflow rate in the inner disk and its long-term average. 

The frequency-luminosity correlation, for individual
sources, seems to depend strongly on a characteristic timescale
associated with $\dot{M}$ variation. Estimation of the source distance and
therefore of the $\dot{M}$ range can be very important in modelling
the \emph{parallel tracks }of a single source. In the case of many
sources, however, each track in the frequency-luminosity plane
corresponds to a different source with its own
intrinsic properties such as its mass, radius, spin, and magnetic
field in addition to its $\dot{M}$ range. For the ensemble of neutron star LMXBs, the huge luminosity differences between sources sharing similar frequency ranges for kHz QPOs can be accounted for if a parameter determining the QPO frequency in addition to $\dot{M}$ differs from one source to another \citep{Mendez2001}. As the QPO frequencies are usually associated with certain characteristic radii at which the neutron star interacts with the accretion flow, it is plausible to consider the stellar magnetic field strength as the relevant parameter beside $\dot{M}$.

In this paper, we search for possible correlations between QPO frequency and accretion-related parameters such as the mass accretion rate, $\dot{M}$, inferred from $L_{\mathrm{X}}$ and $\dot{M}/B^{2}$, where $B$ is the magnetic field strength on the
surface of the neutron star. We find a correlation between the lower kHz QPO frequency,
$\nu_{1}$, and $\dot{M}/B^{2}$. We consider the scaling of $\nu_{1}$ with the Keplerian frequency at the magnetopause and conclude that the new correlation suggests the magnetic boundary region as the origin of lower kHz QPOs.

In Section~\ref{analys}, we describe the analysis and results. Our analysis consists of two parts: (1) estimation of source distances in comparison with the distance values used in
\citet{Ford2000}, and (2) search for a possible correlation between the kHz
QPO frequency and the accretion-related parameters, $\dot{M}$ and $\dot{M}/B^{2}$, based on source luminosities, which we calculate according to the up-to-date source distances. In Section~\ref{disc}, we discuss our results and present our conclusions.

\section{Analysis}\label{analys}

\subsection{Distance Estimation}\label{dist}

We determine the source distances by means of red clump giants (RCGs).
Calculation of the distance to the source depends on the near-infrared (NIR)
extinction, $A_{K_{s}}$, for each source. We use the hydrogen column
densities, $N_{H}$, in the literature and convert them to the extinction in
visual band ($A_{V}$), which in turn can be converted to $A_{K_{s}}$.

\subsubsection{Determination of Extinctions}\label{extinc}

We convert $N_{H}$ values in Table~\ref{table1} to $A_{V}$ values using $%
N_{H}=(2.21\pm 0.09)\times 10^{21}\times A_{V}$ \citep{Guver2009}. The
optical extinctions are then converted to NIR extinctions using $%
A_{K_{s}}=R_{K_{s}}\times A_{V}$, where $R_{K_{s}}$ coefficients such as $%
0.062$, $0.085$, and $0.112$ were found empirically by
\citet{Nishiyama2008}, \citet{Nishiyama2006}, and \citet{Rieke1985},
respectively. We use galactic open clusters to determine the most
precise coefficient.

We select the galactic open clusters near the galactic plane with
well-known distance and reddening given in \citet{Kharchenko2005}
catalogue. Using different $R_{K_{s}}$ coefficients, we calculate
the distances of the open clusters through the RCG method to be
mentioned in the following section. As seen in Figure~\ref{fig1}a,
we find the most accurate coefficient to be $R_{K_{s}}=0.085$
\citep{Nishiyama2006} and calculate NIR extinctions using
$A_{K_{s}}=0.085\times A_{V}$.

\subsubsection{RCGs as a Distance Indicator}\label{rcg}

Due to their very narrow luminosity function, the absolute magnitude of RCGs can be
assumed to be constant \citep{Stanek1998}. Their
absolute magnitude and intrinsic color in NIR bands are well-known
\citep{Alves2000,Yaz2013}. We use
$M_{K_{s}}=-1.595\pm 0.025$ mag and $(J-K_{s})_{0}=0.612\pm 0.003$
mag given by \citet{Yaz2013}.

Using 2MASS data \citep{Cutri2003} in the field of view of the
sources listed in Table~\ref{table1}, we select all the stars within a given
radius and identify RCGs with the help of the Galaxia extinction model %
\citep{Sharma2011,Binney2014}. The extinction in terms of distance and coordinates is
estimated through the 2MASS color-magnitude diagram (CMD) centered around
the source. In Table~\ref{table1}, we employ
the UKIDSS data \citep{Lucas2008} for sources such as GX 17+2 and GX 5--1
whose distances exceed the limiting magnitude of the 2MASS photometry.

\begin{figure}
\epsscale{1.1} \plotone{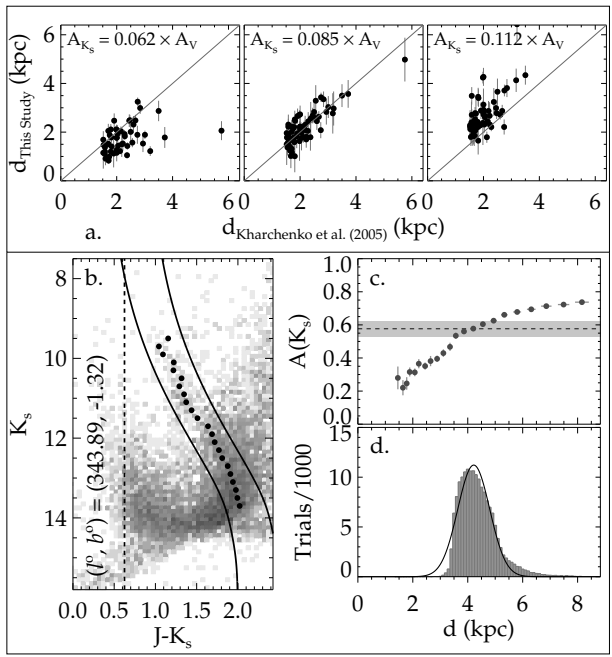} \caption{(a) Comparison of the
calculated distances of the galactic open clusters with those in the
literature using different coefficients. (b) 2MASS CMD centered
around 4U 1608-522 with a radius of $r=10^\prime$. (c)
Extinction-distance relation for the field. The horizontal dashed
line and its gray band show the $K_s$-band extinction value of the
source and its uncertainty, respectively. (d) Results of the MCMC
simulation for the source.\label{fig1}}
\end{figure}

The 2MASS CMD centered around 4U 1608$-$522 with a radius of $r=10^{\prime }$
is shown in Figure~\ref{fig1}b. In this panel, the dashed line represents
the intrinsic color of RCGs while the solid lines indicate the selection of
RCGs using the Galaxia extinction model. We split the RCG data according to $%
K_{s}$ magnitudes in $0.2$ mag bins (dots between solid lines in Figure~\ref%
{fig1}b). We calculate the mean color, $(J-K_{s})$, for each bin. The
extinction can then be determined as
\begin{equation}
A_{K_{s}}=0.528\times \lbrack (J-K_{s})-(J-K_{s})_{0}]  \label{aks}
\end{equation}%
\citep{Nishiyama2009}. We estimate the uncertainty of the mean color for
each bin, $\sigma _{(J-K_{s})}$, and find the total uncertainty of the
extinction, $\sigma _{{A_{K_{s}}}}$, taking into account the uncertainty of the
intrinsic color, $\sigma _{(J-K_{s})_{0}}$.
We determine the distance,
\begin{equation}
d=10^{\left( \mu _{K_{s}}+5\right) /5}\;\mathrm{pc},  \label{distance}
\end{equation}
and its uncertainty using the distance modulus, $\mu
_{K_{s}}=m_{K_{s}}-M_{K_{s}}-A_{K_{s}}$ and its uncertainty,%
\begin{equation}
\sigma _{\mu _{{K_{s}}}}^{2}=\sigma _{m_{{K_{s}}}}^{2}+\sigma
_{M_{K_{s}}}^{2}+\sigma _{{A_{K_{s}}}}^{2},  \label{uncert}
\end{equation}
respectively. Here, $\sigma _{M_{K_{s}}}=0.025$ mag \citep{Yaz2013}
and $\sigma _{m_{{K_{s}}}}$ is the mean value of apparent
magnitude errors in each bin. The $A_{K_{s}}-d$ relation we obtain
for 4U 1608-522 is shown in Figure~\ref{fig1}c \citep{Guver2010}. We use the relation
as a model to calculate the source distance through a Markov chain
Monte Carlo (MCMC) simulation with 200000 trials. We fit a Gaussian
function to the MCMC results to determine the distance and its error
as shown in Figure~\ref{fig1}d. For details of the method, see
\citet{Ford2005}. The estimated distances to 8 LMXB sources are tabulated
in Table~\ref{table1} with the calculated $K_{s}$-band extinction using
the $N_H$ values in the literature. The distances to other LMXB sources
couldn't be estimated because of the large error introduced by nearly
constant extinction at high galactic latitudes and the low extinction
in certain regions of the galactic plane. For these sources, including 4U~1636--53, KS~1731--260, and 4U~1705--44, we adopt either the distance values quoted in \citet{Ford2000} or the recent distance estimates in the literature (Table~\ref{table1}).

\subsection{Correlations with Lower kHz QPO Frequency}\label{nu1cor}

The lack of correlation between the kHz QPO frequencies, $\nu _{\mathrm{kHz}%
} $, and the X-ray luminosity in the ensemble of LMXB sources was
revealed by \citet{Ford2000}. Whether there exists a correlation
that holds between the kHz QPO frequencies and the accretion-related
parameters in the presence of many sources is still an open
question. As a part of our attempt towards understanding the
decoupling between $\nu _{\mathrm{kHz}}$ and $L_{\mathrm{X}}$, we
focus on the possible coupling between $\nu _{\mathrm{kHz}}$ and the
parameters such as $\dot{M}$ and $\dot{M}/B^{2}$, where $B$ is the
surface magnetic field strength of the neutron star accreting mass
at a rate $\dot{M} $.

The number of data points is crucial for the statistical significance of the
correlation. Among all LMXB sources with available data of
$\nu _{\mathrm{kHz}}$ as a function of $L_{\mathrm{X}}$, 4U 1608--52 and Aql X-1
contribute with the highest number of data points to the lower kHz
QPO frequencies whereas the upper kHz QPOs are either weak to be
detected or absent for these two sources \citep{Ford2000}. The lower
kHz QPO peak with frequency $\nu _{1}$ is usually narrower and
stronger than the upper kHz QPO peak \citep{Mendez1998}. As data of
$\nu _{1}$ as a function of $L_{\mathrm{X}}$ are available for all
sources in \citet{Ford2000}, we focus on the correlation between
$\nu _{1}$ and the accretion-related parameters.

In addition to the data of the source 4U~1636--53 in \citet{Ford2000}, we take into account the relatively recent data of the same source \citep{Belloni2007,Sanna2012}. Our analysis also includes the lower kHz QPO data of XTE~J1701--462 with significance level higher than $3\sigma$ \citep{Sanna2010}. This source is the unique example of a neutron star LMXB that underwent a transition from Z source behavior to atoll source behavior at sufficiently low luminosities \citep{Homan2010}. We refer the reader to Section~\ref{disc} for a brief discussion on the possible inclusion of the single data point of a QPO detected during the Z phase of the source with significance less than $3\sigma$ at $\sim 500$~Hz. In a search for a possible correlation between kHz QPO frequencies and accretion-related parameters in the ensemble of sources, we expect XTE J1701--462 to affect the frequency distribution to some extent as it possesses the largest variation in X-ray luminosity among Z and atoll sources.

\begin{deluxetable}{ccccc}
\tablecolumns{5}
\tabletypesize{\small} \tablewidth{0pt} \tablecaption{Distances\label{table1}} \tablewidth{0pt} \tablehead{ \multicolumn{5}{c}{Estimated distances using RCGs} \\
\cline{1-5} \\
\colhead{Source} & \colhead{$N_H~(10^{22}~\text{cm}^{-1})$} &
\colhead{Ref.} & \colhead{$A_{K_s}$} & \colhead{$d$ (kpc)} }
\startdata
4U 1608--522 & $1.5\pm0.1$ & 1 & $0.58\pm0.05$ & $4.2\pm0.6$ \\
4U 1702--42 & $1.95$ & 2 & $0.75\pm0.04$ & $6.1\pm0.7$ \\
4U 1728--34 & $2.61\pm0.07$ & 3 & $1.00\pm0.04$ & $4.3\pm0.2$ \\
4U 1636--53 & $0.32\pm0.01$ & 4 & $0.12\pm0.04$ & $4.7\pm3.4$ \\
KS 1731--260 & $1.06\pm0.08$ & 5 & $0.41\pm0.05$ & $5.4\pm2.9$ \\
4U 1705--44 & $1.42\pm0.06$ & 6 & $0.55\pm0.04$ & $9.1\pm3.2$ \\
GX 17+2 & $2.00\pm0.05$ & 7 & $0.77\pm0.04$ & $9.9\pm0.8$ \\
GX 5--1 & $3.0$ & 8 & $1.2\pm0.1$ & $8.8\pm0.3$ \\
\cutinhead{Adopted distances} 
Source & & Ref. & & $d$ (kpc) \\ 
\tableline \\
4U 1636--53 & ................... & 9 & ................... & $6.0\pm0.5$ \\
KS 1731--260 & ................... & 10 & ................... & 8.5 \\
4U 1705--44 & ................... & 10 & ................... & 11 \\
Aql X-1 & ................... & 10 & ................... & 3.4 \\
4U 1735--44 & ................... & 10 & ................... & 7.1 \\
Sco X-1 & ................... & 10 & ................... & $2.8\pm0.3$ \\
4U 0614+09 & ................... & 11 & ................... & $3.2\pm0.5$ \\
4U 1820--30 & ................... & 12 & ................... & $7.6\pm0.4$ \\
Cyg X-2 & ................... & 13 & ................... & 11.4--15.3 \\
XTE J1701--462 & ................... & 14 & ................... & 8.8 \\
\enddata
\tablecomments{Adopted distances are used in the analysis only for sources whose distances cannot be calculated using the RCG method and the sources such as 4U~1636--53, KS~1731--260, and 4U~1705--44 whose distances could be estimated using the RCG method, but with unreasonably large errors. The error of the adopted distance for 4U~1636--53, however, could be an underestimate as the error is based only on the fluctuation of the high peak flux of the bursts ignoring the possible uncertainties associated with the hydrogen mass abundance and the anisotropy of the emission during bursts. The adopted distance of 4U~1636--53, nonetheless, complies with the earlier estimations of the distance to the same source.}
\tablerefs{(1) \citealt{Penninx1989}; (2) \citealt{Guver2012}; (3)
\citealt{Dai2006}; (4) \citealt{Lyu2014}; (5)
\citealt{Rutledge2002}; (6) \citealt{Di Salvo2005}; (7)
\citealt{Farinelli2005}; (8) \citealt{Jackson2009}; (9) \citealt{Galloway2008}; (10) \citealt{Ford2000}; (11) \citealt{Kuulkers2010}; (12) \citealt{Kuulkers2003}; (13) \citealt{Jonker2004}; (14) \citealt{Lin2009}.}
\end{deluxetable}

\subsubsection{Power-Law Fit to Frequency Distribution}\label{plfit}

We handle the frequency-luminosity data in \citet{Ford2000} incorporated with the data of 4U~1636--53 and XTE~J1701--462 \citep{Belloni2007,Sanna2010} to obtain the current distribution of sources in the $\nu _{1}$ versus $L_{\mathrm{X}}$ plane. We update the luminosity information using
the constancy of $L_{\mathrm{X}}/d^{2}$, for the current value of
the source distance, $d$, with respect to the its value quoted in
\citet{Ford2000}. As the distance value given
by \citet{Ford2000} as 9.5 kpc for the source GX\ 340+0 does not
match the value estimated for the same source (11.8 kpc) by the
reference therein, we exclude this source from the present analysis.
The exclusion of GX 340+0 does not affect our analysis as the number
of data points contributed by this source to the statistical
significance of the correlation is limited to 3 \citep{Ford2000}.

\begin{figure}
\epsscale{1.18} \plotone{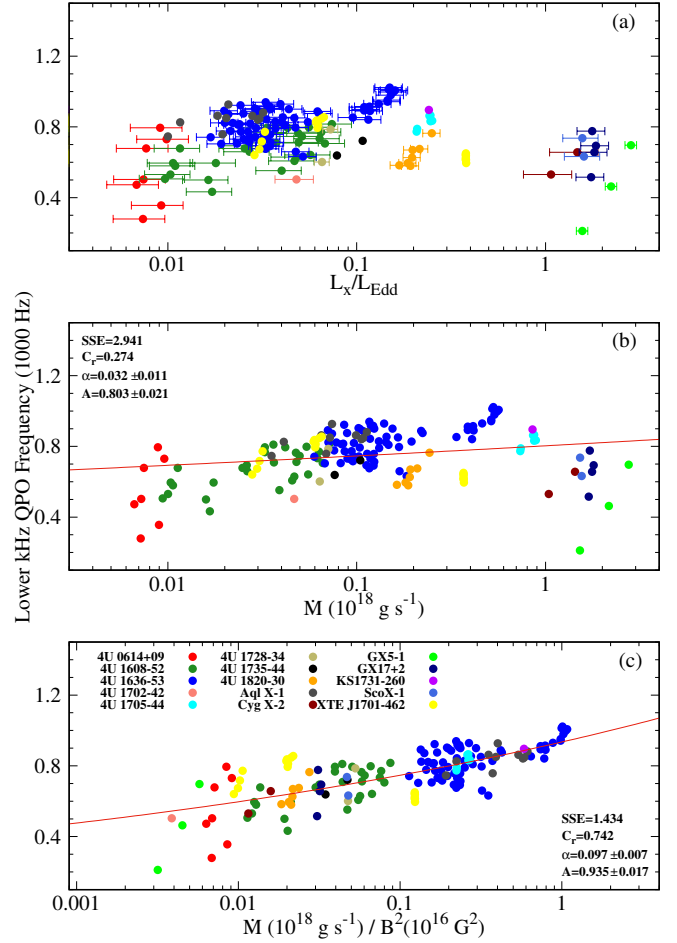} \caption{Distribution of sources in
the plane of lower kHz QPO frequency vs. $L_{\mathrm{X}}$ in the
2--50 keV band reported by Ford et al. (2000) according to the
current distances and their uncertainties (horizontal error bars) estimated by the
present work (panel~a). Lower kHz QPO frequency vs. $\dot{M}$ and $\dot{M}%
/B^{2}$ are shown (without error bars) in panels~b and c, respectively. All panels include the additional data of 4U 1636--53 \citep{Belloni2007} and the data of XTE J1701--462 \citep{Sanna2010}. For each source, $\dot{M}$ is inferred from $L_{\mathrm{X}}$ using the mass and radius estimation by the FPS EoS \citep{PR1989,Cook1994}.\label{fig2}}
\end{figure}

The up-to-date distribution of 15 LMXB sources in the $\nu _{1}$ versus $L_{%
\mathrm{X}}$ plane is shown in Figure~\ref{fig2}a. The range of lower kHz
QPO frequencies is roughly between 200 and 1000 Hz for each source that
differs by orders of magnitude in its X-ray luminosity from any other source
in the same distribution. We confirm the absence of any correlation between $%
\nu _{1}$ and $L_{\mathrm{X}}$, which was first pointed out by
\citet{Ford2000}. To search for a possible correlation between $\nu
_{1}$ and accretion-related parameters, we assume that
$L_{\mathrm{X}}$ is a good indicator of accretion luminosity, that is, $L_{\mathrm{X}}\simeq GM\dot{M}%
/R $, where $\dot{M}$ is the mass accretion rate, $M$ and $R$ are the mass
and radius of the neutron star, respectively. Note that $\dot{M}$ can be
inferred from $L_{\mathrm{X}}$ provided that the compactness ratio, $M/R$,
is known. For the majority of equations of state (EoS) in the literature, $%
M/R\sim 0.5-2$ with $M$ and $R$ being normalized to $1M_{\odot }$
and 10 km, respectively. For illustrative purposes, in Figure~\ref{fig2}b,
we show the distribution of sources in the $\nu _{1}$
versus $\dot{M}$ plane for the best correlation between $\nu _{1}$
and $\dot{M}$ using the FPS EoS in \citet{PR1989}. The distribution
in Figure~\ref{fig2}b is obtained by shifting the $\dot{M}$ range
(inferred from the $L_{\mathrm{X}}$ range) of each source according
to different $M$ and $R$ values (Table~\ref{table2}) estimated by
the static limit of the FPS EoS \citep{Cook1994} until the best correlation with a minimum sum of squared errors $(SSE)$ is
found. We repeat the same procedure for other EoS such as SQM3 \citep{PCL1995}. The source distribution is not expected to be sensitive to the choice of EoS as the compactness ratio has a narrow range of values $(\simeq 0.5-2)$ in the zoo of EoS. The minimization of the sum of squared errors (residuals)
between data and the values of the model function is done using
the Marquardt-Levenberg algorithm. The correlation coefficient,
\begin{equation}
C_{\mathrm{r}}=\sqrt{1-\frac{SSE}{SST}}, \label{corcoef}
\end{equation}
can be used as an estimate of correlation strength \citep{Marquardt1963,MR1987,Press2007}. The value of $C_{\mathrm{r}}$, however, does not yield a direct measure of the correlation strength. To the extent that $C_{\mathrm{r}}$ is different from zero, it is possible to reject the null hypothesis that two variables are uncorrelated. The strongest correlation corresponds to $C_{\mathrm{r}}=1$ ($SSE=0$). We calculate the sum of squared errors using
\begin{equation}
SSE=\sum_{i}{(y_{i}-f_{i})^2}
\end{equation}
and the sum of squared totals using
\begin{equation}
SST=\sum_{i}{(y_{i}-\bar{y})^2},
\end{equation}
where $\bar{y}$ is the mean value of data $y_{i}$ and $f_{i}$ is the value of the fit function at the $x$-coordinate of $y_{i}$.
We model the correlation of $\nu
_{1}$ with $\dot{M}$ through a power-law:
\begin{equation}
\frac{\nu _{1}}{1000\,\mathrm{Hz}}=A\left( \frac{\dot{M}}{10^{18}\,\mathrm{%
g\,s}^{-1}}\right) ^{\alpha }.  \label{nu1mdot}
\end{equation}
The model parameters for the best fit $(SSE=2.941, C_{\mathrm{r}}=0.274)$ to all sources in Figure~\ref{fig2}b are $A=0.803\pm 0.021$ and $\alpha =0.032\pm 0.011$. Note that the distribution of Z and atoll sources weakens the correlation between $\nu
_{1}$ and $\dot{M}$. We find similar results if we use SQM3 EoS in
\citet{PCL1995} to choose $M$ and $R$ values. The power-law parameters for the best fit $(SSE=2.939, C_{\mathrm{r}}=0.275)$ to the source distribution in the $\nu_{1}$ versus $\dot{M}$ plane are $A=0.804\pm 0.021$ and $\alpha =0.035\pm 0.010$ if SQM3 EoS is used. The correlation of $\nu _{1}$ with $\dot{M}$, however, is not conclusive for the entire ensemble of Z and atoll sources as the correlation coefficient $(C_{\mathrm{r}}\simeq 0.27)$ is far from being significantly different from zero and close to one.

The existence of a correlation between kHz QPO frequencies and a fundamental
parameter describing the neutron star interaction with its accretion flow
can be very important towards understanding the physical mechanism behind
these milisecond oscillations. Although the neutron stars in LMXBs are
usually thought to be weakly magnetized with surface field strengths in the $%
10^{8}-10^{9}$ G range, magnetic field may still play an important
role beside mass accretion rate in identifying the relation between
the spectral and timing properties of Z and atoll sources
\citep{HK1989}. The decoupling between $\nu _{\mathrm{1}}$ and $L_{\mathrm{X}}$ or $%
\nu _{\mathrm{1}}$ and $\dot{M}$ in the ensemble of different
sources of different spectral classes might be primarily due to the
difference in the magnetic field strength, $B$, on the surface of
the neutron star in each source. As discussed in the next section,
$\dot{M}/B^{2}$ is expected to be the parameter correlating with
$\nu _{\mathrm{kHz}}$. In Figure~\ref{fig2}c, we display the best
correlation between $\nu _{1}$ and $\dot{M}/B^{2}$ throughout the
entire ensemble of sources using the FPS EoS in \citet{PR1989} as an
example. As summarized in Table~\ref{table2}, we obtain the
distribution of sources in Figure~\ref{fig2}c by calibrating the $B$
value of each source with a possible pair of $M$ and $R$ estimated
by the EoS. We parametrize the correlation of $\nu _{1}$ with
$\dot{M}/B^{2}$ as a power-law:
\begin{equation}
\frac{\nu _{1}}{1000\,\mathrm{Hz}}=A\left[ \frac{\dot{M}/10^{18}\,\mathrm{%
g\,s}^{-1}}{\left( B/10^{8}\,\mathrm{G}\right) ^{2}}\right] ^{\alpha }.
\label{nu_md_bs}
\end{equation}%
We find $A=0.935\pm 0.017$ and $\alpha =0.097\pm 0.007$ for the best fit $(SSE=1.434, C_{\mathrm{r}}=0.742)$ in Figure~\ref{fig2}c. We obtain the best fit $(SSE=1.461, C_{\mathrm{r}}=0.735)$ with $A=0.962\pm 0.019$ and $\alpha =0.100\pm 0.008$ to a similar
distribution using SQM3 EoS in \citet{PCL1995}. Being independent of any equation
of state, we find the best fit $(SSE=1.395, C_{\mathrm{r}}=0.75)$ with $A=0.930\pm 0.016$ and $\alpha =0.094\pm 0.007$. The mass and radius estimations for a distribution like
that in Figure~\ref{fig2}c are subject to uncertainties arising from
the choice of magnetic field strength. Similarly, any uncertainty
associated with distance and therefore luminosity estimation can be
incorporated within the uncertainty of $B$ value that we assign to
each source to end up with the same distribution. We expect the
correlation between $\nu _{1}$ and $\dot{M}/B^{2}$ to hold in the
current ensemble of Z and atoll sources according to Equation
(\ref{nu_md_bs}) more or less with $A\approx 1$, $\alpha \approx
0.1$, and $C_{\mathrm{r}}\simeq 0.75$. The correlation between $\nu_{1}$ and $\dot{M}/B^2$ is much more pronounced than that of $\nu_{1}$ with $\dot{M}$ as indicated by the closeness of $C_{\mathrm{r}}$ to one.

\begin{deluxetable}{lcccccc}
\tablecolumns{7} \tablewidth{0pc} \tablecaption{Correlation-based
estimations of mass, radius, and magnetic field.\label{table2}}
\tablehead{ \colhead{}    & \multicolumn{2}{c}{$\nu _{1}$ vs.
$\dot{M}$} & \colhead{}   &
\multicolumn{3}{c}{$\nu_{1}$ vs. $\dot{M}/B^{2}$} \\
\cline{2-3} \cline{5-7} \\
\colhead{Source} & \colhead{$M/M_{\odot}$}   & \colhead{$R$ (km)} &
\colhead{}    & \colhead{$M/M_{\odot}$}   & \colhead{$R$ (km)} &
\colhead{$B$ ($10^{8}$ G)}} \startdata
4U 0614+09 & 1.8 & 9.35 &  & 1.7 & 10.2 & 1.1 \\
4U 1608--52 & 1.8 & 9.35 &  & 1.4 & 10.8 & 1.1 \\
4U 1636--53 & 0.6 & 11.4 &  & 1.06 & 11.1 & 0.54 \\
4U 1702--42 & 1.8 & 9.35 &  & 1.8 & 9.5 & 3.5 \\
4U 1705--44 & 1.8 & 9.35 &  & 1.3 & 11 & 1.2 \\
4U 1728--34 & 1.8 & 9.35 &  & 1.4 & 10.8 & 1.4 \\
4U 1735--44 & 1.8 & 9.35 &  & 1.4 & 10.8 & 1.8 \\
4U 1820--30 & 1.8 & 9.35 &  & 1.4 & 10.8 & 3.6 \\
Aql X-1 & 0.6 & 11.4 &  & 1.2 & 11 & 0.3 \\
Cyg X-2 & 1.8 & 9.35 &  & 1.4 & 10.8 & 11.5 \\
GX 5--1 & 1.8 & 9.35 &  & 1.8 & 9.35 & 21.9 \\
GX 17+2 & 1.8 & 9.35 &  & 1.4 & 10.8 & 9.0 \\
KS 1731--260 & 0.6 & 11.4 &  & 1.3 & 11 & 0.8 \\
Sco X-1 & 1.8 & 9.35 &  & 1.4 & 10.8 & 6.9 \\
XTE J1701--462 & 1.8 & 9.35 &  & 1.53 & 10.7 & 2.0 
\enddata
\tablecomments{Mass, radius, and magnetic field values for sources
in the $\nu_1$ vs. $\dot{M}$ and $\nu_1$ vs. $\dot{M} /B^{2}$ planes
are estimated using the FPS EoS \citep{PR1989,Cook1994}.}
\end{deluxetable}

Although it fairly describes the average run of $\nu_{1}$ over the ensemble of sources, the slope of the power-law fit line in Figure~\ref{fig2}c, however, cannot account for the distribution of QPO frequencies exhibited by an individual source. Different tracks (the so-called \emph{parallel tracks}) with different slopes are followed by each source (Figure~\ref{fig2}). This is the main reason why sources in the $\nu_{1}$ versus $\dot{M}/B^2$ plane are scattered all along the power-law fit line and the strength of correlation is limited to $C_{\mathrm{r}}\simeq 0.75$. If there exists a relation between the lower kHz QPO frequency, $\nu_{1}$ and the accretion-related parameter, $\dot{M}/B^2$, then it cannot be described by a simple power-law in the case of individual sources. Indeed, $\dot{M}/B^2$ becomes simply $\dot{M}$ for an individual source once the magnetic field, $B$, is fixed. In the presence of many sources, however, $B$, in combination with $\dot{M}$, can be crucial for understanding the QPO frequency distribution in the ensemble of sources.

Apart from the observed scattering of sources for which we come up with an explanation in Section~\ref{mffit}, the results of the present analysis reveal the existence of a possible correlation of the lower kHz QPO frequencies in the ensemble of Z and atoll
sources with the accretion-related parameter $\dot{M}/B^{2}$.
The reason behind the absence of correlation between
$\nu _{\mathrm{kHz}}$ and $L_{\mathrm{X}}$ in the same
population can also be realized if the $\nu _{\mathrm{kHz}}$ range
is set by a physical process working at a particular length scale,
which is not only determined by $\dot{M}$. The relevant length scale
imposed by an accreting neutron star interacting with the accretion
disk through its magnetosphere can be estimated by the Alfv\'{e}n radius,%
\begin{equation}
r_{\mathrm{A}}\simeq \left( \frac{R^{12}}{GM}\right) ^{1/7}\left( \frac{\dot{%
M}}{B^{2}}\right) ^{-2/7},  \label{ralfv}
\end{equation}
where $B$ represents the magnetic dipole field strength on the surface of
the neutron star \citep{Cui2000}. In the innermost regions of an accretion disk interacting
with the neutron star magnetosphere, all dynamical frequencies can be
measured with respect to the Keplerian frequency at the Alfv\'{e}n radius,%
\begin{equation}
\nu _{\mathrm{A}}\equiv \nu _{\mathrm{K}}(r_{\mathrm{A}})=\nu _{0}\left(
M,R\right) \left( \frac{\dot{m}}{b^{2}}\right) ^{3/7},  \label{nu_kep}
\end{equation}%
where $\dot{m}\equiv \dot{M}/10^{18}\,\mathrm{g\,s}^{-1}$ and $b\equiv
B/10^{8}\,\mathrm{G}$ are the dimensionless mass inflow rate and field
strength, respectively. We write the mass and radius dependence of $\nu _{%
\mathrm{A}}$ as%
\begin{equation}
\nu _{0}\left( M,R\right) \simeq 1955\,\mathrm{Hz\,}\left( \frac{M}{M_{\odot
}}\right) ^{5/7}\left( \frac{R}{10\,\mathrm{km}}\right) ^{-18/7}.
\label{nu_zr}
\end{equation}%
It is plausible, based on their observational properties, that the kHz QPO
frequencies are scaled by the dynamical frequencies in the inner disk and
thus by $\nu _{\mathrm{A}}$. Two different sources of similar neutron star
masses and radii but different field strengths, e.g., $b_{1}=1$ and $b_{2}=10
$, can exhibit the kHz QPOs of similar frequencies for the accretion rates $%
\dot{m}_{1}=0.01$ and $\dot{m}_{2}=1$ (Equation~\ref{nu_kep}). If $\dot{M}$
is a good indicator of $L_{\mathrm{X}}$ and the QPO frequencies are mainly
determined by the magnetospheric length scale, a similar range of $\nu _{%
\mathrm{kHz}}$ may then hold for sources differing by orders of magnitude in
$L_{\mathrm{X}}$ while differing by at most one order of magnitude in $B$.

Based on the assumption that all neutron stars in LMXBs are close to spin equilibrium, \citet{WZ1997} proposed $L_{\mathrm{X}} \propto B^{2}$ as a relation between X-ray luminosity and surface magnetic field strength in accordance with the early interpretation that the frequency difference between upper and lower kHz QPO frequencies is a good indicator of neutron star spin frequency. Later studies such as those by \citet{MB2007} and \citet{Altamirano2010}, however, pointed out that this interpretation may not be correct at all. According to \citet{WZ1997}, almost all sources share a narrow range of spin period, which suggests similar values for magnetospheric radii and therefore nearly constant $L_{\mathrm{X}}/B^{2}$. Although their idea is also based on magnetosphere-disk interaction, it is considerably different from what we propose here. It is plausible that sources having similar magnetospheric length scales could produce similar range of QPO frequencies in the absence of any other length scale or boundary condition (see, e.g., the following section for the effect of boundary region width on the range of QPO frequencies). However, $\dot{M}/B^2$ should span more than 2 orders of magnitude range (Figure~\ref{fig2}c) in the present analysis and therefore cannot be treated as a constant throughout the population of neutron star LMXBs in order to explain the observed range of kHz QPO frequencies.

The scaling of kHz QPO frequencies can be done using $\nu _{\mathrm{A}}$ if
these high frequency QPOs are magnetospheric in origin. The model proposed by \citet{Zhang2004} associates the upper kHz QPO frequency with the Keplerian frequency at the preferred radius ($\simeq r_{\mathrm{A}}$), i.e., with $\nu_{\mathrm{A}}$ in Equation~(\ref{nu_kep}). The lower kHz QPO frequency can then be estimated as $\nu_{1}\propto \nu_{\mathrm{A}}^2$ \citep{Zhang2004}, which implies $\alpha=6/7$ as the power-law index of the correlation in Equation~(\ref{nu_md_bs}). Regarding the ensemble of neutron star LMXBs, the present correlation  between $\nu _{1}$ and $\dot{M}/B^{2}$ with
$\alpha \approx 0.1$ cannot be satisfied, however, if $\nu _{1}$ is directly
estimated by either $\nu _{\mathrm{A}}$ or $\nu _{\mathrm{A}}^2$ (Equations \ref{nu_md_bs} and \ref{nu_kep}). Within the context of power-law fit to frequency distribution, it is not clear how to explain the $\dot{M}/B^{2}$ dependence of $\nu _{1}$ revealed by the present correlation that holds for the ensemble of sources. In a rather different context, however, working out the scaling of $\nu _{1}$ with $\nu _{\mathrm{A}}$ can be useful towards a better understanding of the origin of these milisecond oscillations (see Section~\ref{mffit}). The frequency distribution in the case of individual sources as well as the ensemble of sources could then be explained in terms of well-defined physical length and time scales.

\subsubsection{Model Function Fit to Frequency Distribution}\label{mffit}

The oscillatory modes in the innermost regions of accretion disks
were shown to be a likely source of high-frequency QPOs from neutron
star and black hole sources in LMXBs
\citep{AP2008,EPA2008,Erkut2011}. In a non-Keplerian boundary region
near the magnetopause between the neutron star and the disk, the
fastest growing mode frequencies are given by the angular frequency
bands of $\kappa $ and $\kappa \pm \Omega $, where the orbital
angular frequency of the innermost disk matter, $\Omega $, is less
than its Keplerian value $\Omega _{\mathrm{K}}$ and the radial
epicyclic frequency,
\begin{equation}
\kappa =\Omega \sqrt{4+2\frac{d\ln \Omega }{d\ln r}},  \label{kappa}
\end{equation}%
can be seen to be the highest dynamical frequency in the inner disk
\citep{AP2008,EPA2008}. According to the boundary region model, the
lower kHz QPO frequency can be estimated by either $2\pi \nu
_{1}=\kappa -\Omega $ or $2\pi \nu _{1}=\kappa $ with $\Delta \nu
=\Omega /2\pi $ being the frequency difference between lower and
upper kHz QPO peaks. The scaling
of $\nu _{1}$ with $\nu _{\mathrm{A}}$ can then be realized using Equation (%
\ref{kappa}). Regardless of the details about the radial profile of $\Omega $
throughout the boundary region of radial width $\delta r_{\mathrm{A}}$, the
radial epicyclic frequency at the Alfv\'{e}n radius can be roughly estimated
as%
\begin{equation}
\kappa (r_{\mathrm{A}})\simeq 2\pi \Delta \nu \sqrt{4+\frac{2}{\delta }%
\left( \frac{\nu _{\mathrm{A}}}{\Delta \nu }-1\right) }  \label{kapa_alfv}
\end{equation}%
using Equation (\ref{kappa}) with $\Omega =2\pi \Delta \nu $ and $d\Omega
/dr\simeq (\Omega _{\mathrm{K}}-\Omega )/\delta r_{\mathrm{A}}$ at $r=r_{%
\mathrm{A}}$. As a dimensionless parameter, $\delta $ characterizes the
radial extension of the boundary region $\left( \delta <1\right) $. Using $%
2\pi \nu _{1}=\kappa -\Omega $ (albeit toward similar conclusions, one could
also choose $\nu _{1}=\kappa /2\pi $), it follows from Equation (\ref%
{kapa_alfv}) that%
\begin{equation}
\nu _{1}=F\left( \nu _{\mathrm{A}},\delta, \Delta \nu \right) \sqrt{\nu _{\mathrm{A}}\Delta \nu }%
,  \label{lkhz}
\end{equation}%
where%
\begin{equation}
F\left(\nu _{\mathrm{A}},\delta,\Delta \nu \right) =\sqrt{\frac{2}{\delta }}\left[ \sqrt{%
1-\left( 1-2\delta \right) \frac{\Delta \nu }{\nu _{\mathrm{A}}}}-\sqrt{%
\frac{\delta \Delta \nu }{2\nu _{\mathrm{A}}}}\right]   \label{f}
\end{equation}%
can be seen to be a slowly varying function when $\Delta \nu /\nu _{\mathrm{A%
}}$ becomes small enough for sufficiently high values of $\dot{m}/b^{2}$
(Equation \ref{nu_kep}). Then, $\nu _{1}\propto \nu _{\mathrm{A}}^{1/2}$
describes the approximate scaling of $\nu _{1}$ with $\nu _{\mathrm{A}}$
(Equation \ref{lkhz}). We write the model dependence of $\nu _{1}$ on the
accretion-related parameter, $\dot{m}/b^{2}$, using Equations (\ref{nu_kep}%
), (\ref{lkhz}), and (\ref{f}), as%
\begin{equation}
\nu _{1}=F\left(\nu_{0},\dot{m}/b^{2},\delta,\Delta \nu \right) \sqrt{\nu _{0}\Delta \nu }\left( \frac{%
\dot{m}}{b^{2}}\right) ^{3/14}.  \label{nu1_md}
\end{equation}
According to Equation (\ref{nu1_md}), the lower kHz QPO frequency behaves as a power-law function of $\dot{m}/b^2$ only for relatively high values of $\dot{m}/b^2$. For $\dot{m}/b^{2}\gtrsim 1$ and $\Delta \nu \simeq 300$~Hz, we obtain $\Delta \nu /\nu_{\mathrm{A}}\lesssim 0.1$ and $F\simeq \sqrt{2/\delta}$ (see Equations~\ref{nu_kep}, \ref{nu_zr}, and \ref{f}). Only then, $F\sqrt{\nu _{0}\Delta \nu }$ in Equation~(\ref{nu1_md}), mimics the constant correlation parameter $A$ in Equation~(\ref{nu_md_bs}) unless $\delta$ and $\Delta \nu$ depend on mass acretion rate. Otherwise, the behavior of $\nu_{1}$ cannot be described by a simple power-law. It is highly likely, however, that both $\delta$ and $\Delta \nu$ vary with $\dot{M}$. The model function also depends on the mass and radius of the neutron star through $\nu_{0}$ (Equations~\ref{nu_kep} and \ref{nu_zr}). For sufficiently high values of $\dot{m}/b^2$, the power-law index estimated by the model function is $3/14\simeq 0.2$, which is much closer than $\alpha=6/7$ \citep{Zhang2004} to the power-law index $(\alpha \approx 0.1)$ of the correlation between $\nu_{1}$ and $\dot{M}/B^2$ in the ensemble of sources (Section~\ref{plfit}). Nevertheless, the power-law behavior of the model function for sufficiently high values of $\dot{m}/b^2$ cannot account for the distribution of sources in the $\nu_{1}$ versus $\dot{M}/B^2$ plane (Figure~\ref{fig2}c), where $\dot{m}/b^2$ covers more than 2 orders of magnitude range including very low values such as 0.01 for which $\nu_{\mathrm{A}}\simeq 270$~Hz becomes comparable in magnitude with $\Delta \nu$. Next, we disclose the overall behavior of the model function in Equation~(\ref{nu1_md}) for a wide range of $\dot{m}/b^2$ rather than for a limited range of this parameter.

The simple power-law description of the frequency distribution in an individual source is not possible due to the presence of different tracks with different slopes. An elaborate study of the frequency distribution in the case of an individual source as well as the ensemble of sources in the $\nu _{1}$ versus $\dot{M}/B^{2}$ plane according to the model estimation (Equation~\ref{nu1_md}) can be performed by a scan through possible values of the model parameters $M$, $R$, $\delta $, and $\Delta \nu $. The parameter $\delta $ may attain
different values in the $\sim 0.01-0.6$ range for the width of a
typical boundary region \citep{EA2004}. The frequency difference,
$\Delta \nu $, may change from one source to another in the $\sim
200-400$ Hz range \citep{Altamirano2010}. The fluctuations in some
of these parameters might be the reason for the observed scattering
of individual sources around the fit curve (Figure~\ref{fig2}c) in
accordance with the model estimation (Equation~\ref{nu1_md}). The
relatively large amount of scattering for 4U 0614+09 in comparison
with Aql X-1 (Figure~\ref{fig2}c) could be, at least partially,
due to the larger variation observed in its $\Delta \nu $ with the
$\sim 230-400$~Hz range while the fluctuation in $\Delta \nu $ for
Aql X-1 is limited to the $\sim 260-290$~Hz range \citep{Altamirano2010}.

\begin{figure}
\epsscale{1.15} \plotone{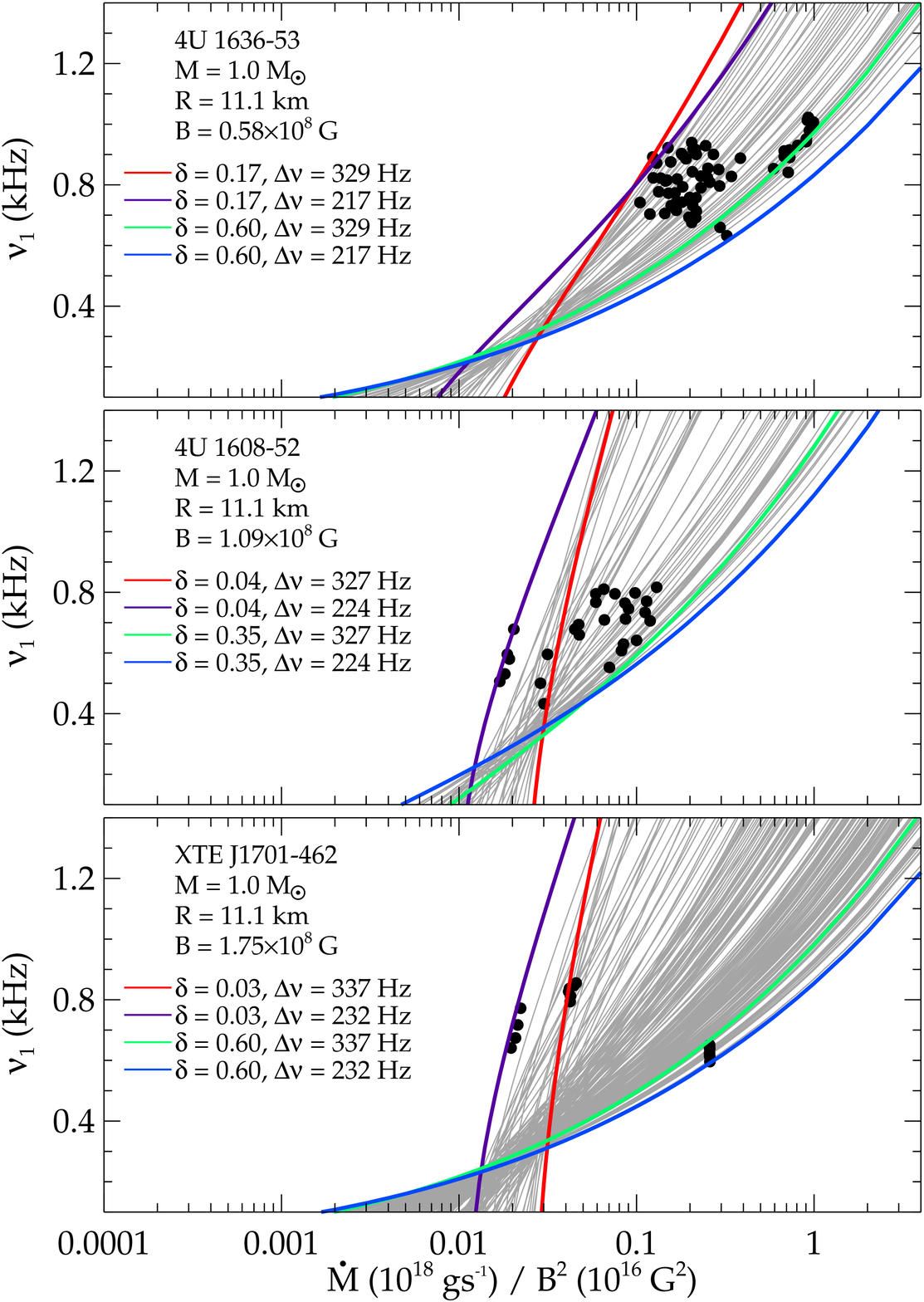} \caption{Model function fit to frequency distribution of individual sources such as 4U 1636--53 (upper panel), 4U 1608--52 (middle panel), and XTE J1701--462 (lower panel) in the plane of lower kHz QPO frequency, $\nu_{1}$, vs. $\dot{M}/B^{2}$ assuming $M/M_{\odot }=1$ and $R=11$~km for the mass and radius of the neutron star, respectively. The magnetic field value for each source is estimated such that the model function, given $M$ and $R$, can be fitted to the observed frequencies through a family of curves, each of which is labeled by a different pair of the model parameters $\delta$ and $\Delta \nu$. For each pair, the model function defines a one-to-one relation between $\nu_{1}$ and $\dot{M}/B^2$. The maximum and minimum values of $\Delta \nu$, are extracted from those of the frequency difference between upper and lower kHz QPO frequencies observed for each source \citep{Yin2007,Sanna2010}. The maximum and minimum values of $\delta$ determine the narrowest possible range that would cover the data of each source. \label{fig3}}
\end{figure}

\begin{figure}
\epsscale{1.18} \plotone{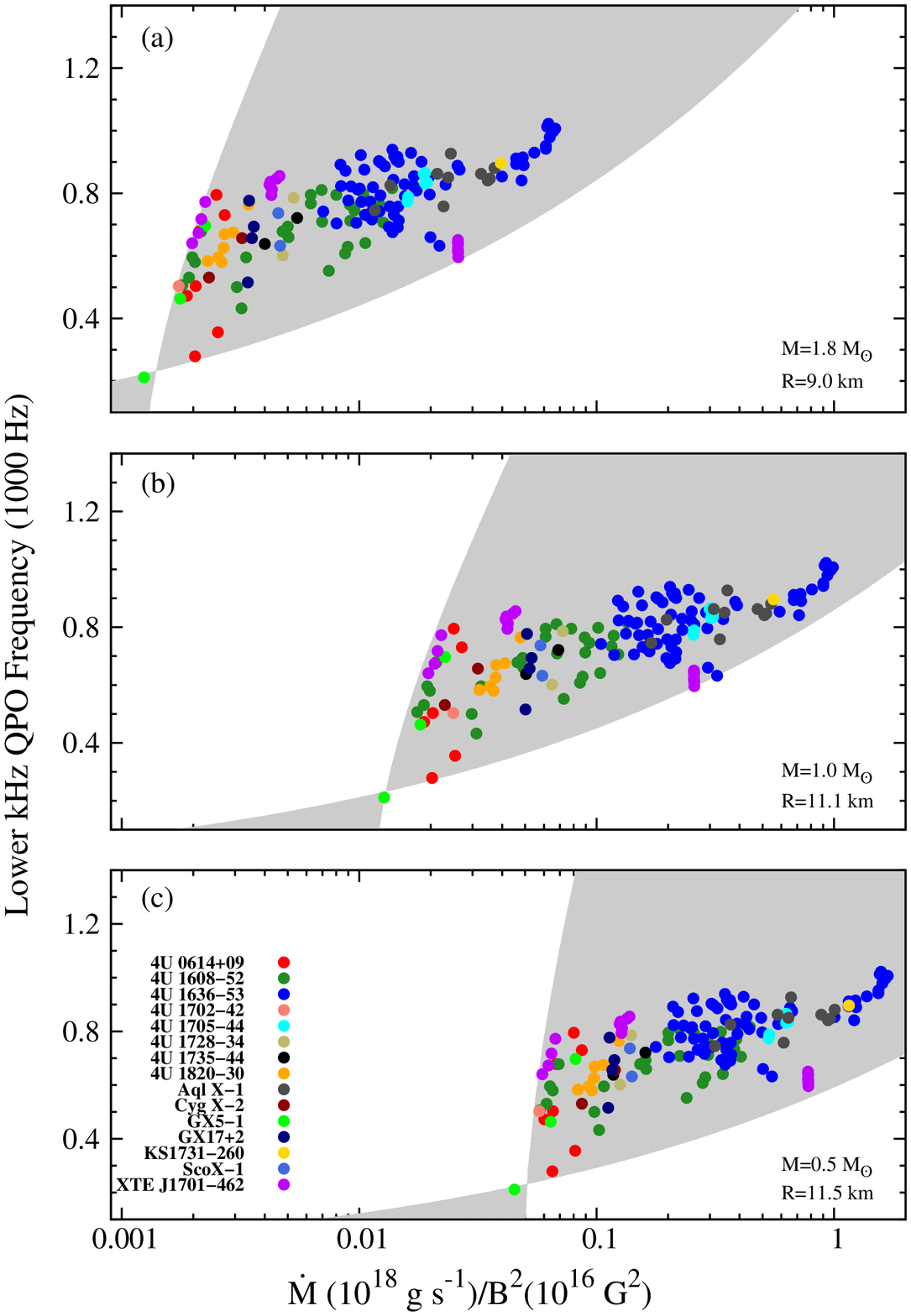} \caption{Estimate of the model function for the distribution of sources in the plane of lower kHz QPO frequency vs. $\dot{M}/B^{2}$ when different values \citep{PR1989,Cook1994} are used for the mass and radius of the neutron star such as $M=1.8M_{\odot }$, $R=9$ km (panel~a), $M=1.0M_{\odot }$, $R=11.1$ km (panel~b), and $M=0.5M_{\odot }$, $R=11.5$ km (panel~c). The magnetic field value for each source (Table~\ref{table3}) is estimated using the same method described in the caption of Figure~\ref{fig3} (see also text for further explanation). In each panel, the shaded region depicts the model function fit to the data of all sources in the ensemble. The ranges of the model parameter, $\delta$, are 0.01--0.52 for $M=0.5M_{\odot}$ (panel~c) and 0.03--0.6 for $M=1.8M_{\odot}$ (panel~a) and $M=1.0M_{\odot}$ (panel~b). \label{fig4}}
\end{figure}

In Figure~\ref{fig3}, we address the
effects of variations in the model parameters such as $\delta$ and $\Delta \nu$ on the distribution of frequencies in the $\nu _{1}$ versus $\dot{M}/B^{2}$ plane. The model function fit to the data of individual sources such as 4U~1636--53, 4U~1608--52, and XTE~J1701--462 in Figure~\ref{fig3} consists of a set of curves, each of which represents a one-to-one relation between $\nu_{1}$ and $\dot{M}/B^2$ when $\delta$ and $\Delta \nu$ are kept fixed (see Equation~\ref{nu1_md}). In each panel of Figure~\ref{fig3}, we determine the magnetic field value of the source through horizontal shifting of the source data in the given plane until the model function fit to them is achieved with the use of the narrowest possible range of $\delta$. We carry out this procedure by assuming $M/M_{\odot }=1$ and $R=11$~km for the mass and radius of the neutron star in each source taking into account, however, the observed range of $\Delta \nu$ for that particular source. Slopes of different data tracks could be accounted for by the gradually varying slope of the model function as $\delta$ and $\Delta \nu$ change within the estimated and observed values of these parameters, respectively (Figure~\ref{fig3}). We note that the smaller the parameter $\delta$ is, the bigger the slope of the model function is. The size of the region where we fit the model function to data is mainly determined by the range of $\delta$. Among 15 LMXB sources, XTE~J1701--462 has the largest variation in $\dot{M}$, which in turn requires the widest range for $\delta$. We find $0.03\leq \delta \leq 0.6$ for XTE~J1701--462 whereas $0.17\leq \delta \leq 0.6$ and $0.04\leq \delta \leq 0.35$ for 4U~1636--53 and 4U~1608--52, respectively.

\begin{deluxetable}{lcccccc}
\tablecolumns{4} \tablewidth{0pt} \tablecaption{Magnetic field estimations for different mass-radius pairs.\label{table3}}
\tablehead{ \colhead{} &
\multicolumn{3}{c}{$B$ ($10^{8}$ G)} \\
\cline{2-4} \\
\colhead{Source} & \colhead{$M,R=1.8,0.9$} & \colhead{$M,R=1.0,1.11$} & \colhead{$M,R=0.5,1.15$}} \startdata
4U 0614+09 & 1.85 & 0.87 & 0.7 \\
4U 1608--52 & 2.25 & 1.09 & 0.85 \\
4U 1636--53 & 1.5 & 0.58 & 0.64 \\
4U 1702--42 & 5.1 & 2.0 & 1.9 \\
4U 1705--44 & 3.5 & 1.3 & 1.3 \\
4U 1728--34 & 3.6 & 1.45 & 1.5 \\
4U 1735--44 & 4.3 & 1.8 & 1.7 \\
4U 1820--30 & 8.3 & 3.3 & 2.95 \\
Aql X-1 & 0.9 & 0.35 & 0.37 \\
Cyg X-2 & 20.8 & 9.85 & 7.3 \\
GX 5--1 & 34.5 & 16.0 & 12.25 \\
GX 17+2 & 22.0 & 8.5 & 8.2 \\
KS 1731--260 & 2.4 & 0.95 & 0.95 \\
Sco X-1 & 18.0 & 7.5 & 7.0 \\
XTE J1701--462 & 3.7 & 1.75 & 1.45 
\enddata
\tablecomments{Magnetic field values in each column are estimated for a mass-radius pair used to obtain the source distributions in Figure~\ref{fig4}. In column headers, $M$ and $R$ represent the normalized mass and radius of the neutron star, i.e., $M/M_{\odot}$ and $R/10$ km, respectively.}
\end{deluxetable}

As seen from Figure~\ref{fig3}, QPO frequencies in sources with similar neutron-star masses and radii can be reproduced within a certain range of $\dot{M}/B^2$ for a given range of $\delta$. Using the broadest range of $\delta$ determined by the peculiar source XTE~J1701--462, it is then possible to incorporate the regions of model function fit to data for different sources and obtain the frequency distribution for the ensemble of sources as in Figure~\ref{fig4}b. Together with the regions of fit for other sources in the ensemble, the superposition of all panels in Figure~\ref{fig3} would lead to such a distribution. We present the results of our analysis for different neutron-star masses and radii such as $M/M_{\odot }=1.8$, $R=9$~km and $M/M_{\odot }=0.5$, $R=11.5$~km in Figures~\ref{fig4}a and \ref{fig4}c, respectively. The shaded region of fit to data in each panel is bounded by the lower and upper limits of $\delta$, which also define the range of the model parameter for XTE~J1701--462. In all panels of Figure~\ref{fig4}, the regions of the model function fit to the data of individual sources are all combined within the shaded region to yield the distribution of sources in the $\nu_{1}$ versus $\dot{M}/B^2$ plane. In the ensemble of sources, the frequency distribution seems to preserve its characteristic shape while the shaded region drifts from higher to lower values of $\dot{M}/B^2$ (i.e., from panel~c to panel~a in Figure~\ref{fig4}) as the neutron-star mass $M$ (radius $R$) increases (decreases) according to Equations~(\ref{nu_zr}) and (\ref{nu1_md}). In order to comprise the data of an individual source within its own region of fit and thus the whole data of the ensemble within the shaded region for a given mass-radius pair, we shift data along the $\dot{M}/B^2$ axis by calibrating the magnetic field of each source. In Table~\ref{table3}, we summarize our estimates of dipole field strength on the surface of the neutron star in each source for different values of mass and radius. Given $M$ and $R$, $B$ cannot have arbitrary values. We can fit the model function in Equation~(\ref{nu1_md}) to individual source data only within a certain range of $\dot{M}/B^2$ even if the largest possible range for $\delta$ is chosen as in the case of XTE~J1701--462. In addition to our concern for fitting the data, shifting procedure also takes into account the relative order of different radii such as the neutron-star radius, $R$, the radius of the innermost stable circular orbit, $r_{\mathrm{ISCO}}$, and the Alfv\'{e}n radius, $r_{\mathrm{A}}$. The frequency distributions in Figure~\ref{fig4} satisfy the condition that $r_{\mathrm{A}}$ is greater than both $R$ and $r_{\mathrm{ISCO}}$ for each source. We note that the distributions in Figure~\ref{fig4}, are similar in shape to the source distribution in Figure~\ref{fig2}c. In particular, the similarity between the distributions in Figures~\ref{fig2}c and \ref{fig4}b is remarkable not only because of shape but also the range spanned by $\dot{M}/B^2$. The reason for this close resemblance is that the range of $\dot{M}/B^2$ throughout which we can fit the model function in Equation~(\ref{nu1_md}) to data depends on the mass and radius of the neutron star with radius dependence being the strongest (see Equations~\ref{nu_zr} and \ref{nu1_md}). The average values of neutron-star mass and radius regarding the distribution in Figure~\ref{fig2}c are $M\simeq 1.4M_{\odot }$ and $R\simeq 10.6$~km, respectively (see Table~\ref{table2}). These values are quite close to $M/M_{\odot }=1$ and $R=11$~km we use to obtain the distribution in Figure~\ref{fig4}b.

\begin{figure}
\epsscale{1.15} \plotone{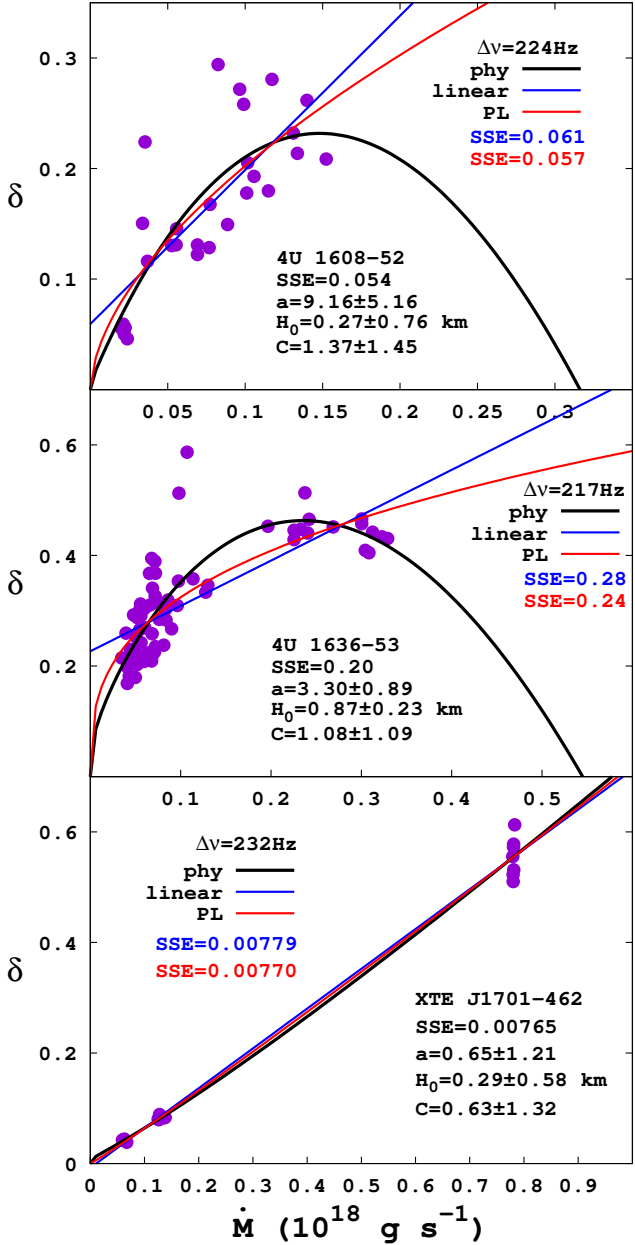} \caption{Variation of the radial width of the boundary region, $\delta$, in units of the innermost disk radius with the mass accretion rate, $\dot{M}$, for different sources. The values of $\delta$ are obtained using the model function in Equation~(\ref{nu1_md}) with $\Delta \nu$ being kept fixed at its minimum value to cover the data of an individual source. In each panel, the black solid curve represents the physically motivated model we employ for $\delta$ as a function of $\dot{M}$, whereas the blue and red solid curves stand for the simple linear and power-law fits to the data, respectively. \label{fig5}}
\end{figure}

The distribution of kHz QPO frequencies in an individual source can be accounted for in terms of variations in the model parameters, $\delta$ and $\Delta \nu$. The boundaries of the region for the model function fit to data are determined by $\Delta \nu$ for a given range of $\delta$ (see Figure~\ref{fig3}). Next, we keep $\Delta \nu$ fixed at the minimum value to ensure the broadest possible zone that can embrace the data of individual sources such as those in Figure~\ref{fig3}. Assuming the same values for the mass, radius, and magnetic field in Figure~\ref{fig3} (see the third column in Table~\ref{table3} for the magnetic field values inferred for all such sources), we find the value of $\delta$ for which the model function can fit to the observed kHz QPO frequency and read the corresponding value of $\dot{M}$ for the given magnetic field strength. In Figure~\ref{fig5}, we display the empirical relation between $\delta$ and $\dot{M}$ for 4U~1608--52, 4U~1636--53, and XTE~J1701--462. In general, $\delta$ can be seen to increase with $\dot{M}$. We observe the similar trend in other sources. In each source, we fit a linear, a power-law, and a physically motivated function to the relation between $\delta$ and $\dot{M}$. The physically motivated expression for $\delta$ as a function of $\dot{M}$ can be derived by assuming that $\delta \propto H_{t}/r_{\mathrm{A}}$. Here, $H_{t}$ represents the typical half-thickness of the accretion disk, which is expected to be in correlation with the half-thickness of the standard Keplerian disk near the magnetopause, that is, $H_{t}=aH_{\mathrm{SS}}(r_{\mathrm{A}})+H_{0}$. Here, $a$ and $H_{0}$ are constants, which are supposed to change from one source to another. Note that the $\dot{M}$ dependence of $\delta$ comes from both $r_{\mathrm{A}}$ and $H_{t}$ as $H_{t}$ also depends on $\dot{M}$ through the half-thickness of the standard disk at the innermost disk radius, i.e., $H_{\mathrm{SS}}(r_{\mathrm{A}})$. The radial width of the boundary region must therefore be a function of mass accretion rate, that is, $\delta$ must vary with $\dot{M}$ simply because the magnetospheric radius is expected to change in response to modifications in $\dot{M}$.

To reveal the dependence of the vertical scale-height in the inner disk on mass accretion rate, we write the half-thickness of the Shakura-Sunyaev disk at the magnetospheric radius as
\begin{equation}
H_{\mathrm{SS}}(r_{\mathrm{A}})\simeq 16 \, \mathrm{km} \,\, \dot{m} \left [1-C\left (\frac{r_{\mathrm{ISCO}}}{r_{\mathrm{A}}}\right )^{1/2}\right ], \label{Hss}
\end{equation}
where $C$ is angular momentum efficiency constant of order unity \citep{SS73}. We obtain the numerical values for the constants of the physically motivated function such as $a$, $H_{0}$, and $C$ as the outcome of the best fit to the relation between $\delta$ and $\dot{M}$ (Figure~\ref{fig5}). In comparison with the simple linear and power-law fits, the description of the data by the model function is better in particular when the number of data points is sufficiently high as in the case of 4U~1636--53 (see the middle panel of Figure~\ref{fig5}). For all other sources including 4U~1608--52 and XTE~J1701--462, the relation between $\delta$ and $\dot{M}$ can be described equally well by the linear, power-law, and physically motivated functions as the number of data points is not high enough to distinguish between different models.

\section{Discussion and Conclusions}\label{disc}

The lack of correlation between kHz QPO frequencies and source luminosities in the ensemble of neutron star LMXBs is the main motivation of our present search to reveal the existence of a possible link between the lower kHz QPO frequency, $\nu_{1}$ and the accretion-related parameter, $\dot{M}/B^2$. The frequency distribution in the plane of $\nu_{1}$ versus mass accretion rate, $\dot{M}$, cannot account for a correlation between these two quantities in the presence of 15 LMXB sources (Figure~\ref{fig2}b). On the other hand, the lower kHz QPO frequency seems to be correlated with $\dot{M}/B^2$ in the ensemble of LMXBs if different neutron-star sources have similar radii. The comparison of Figure~\ref{fig2}c and Table~\ref{table2} with Figure~\ref{fig4} and Table~\ref{table3} is a clear evidence of the fact that the correlation between $\nu_{1}$ and $\dot{M}/B^2$ is a result of the cumulative effect of the model function fit to individual source data (Figure~\ref{fig3}).

The existence of a possible correlation between $\nu_{1}$ and $\dot{M}/B^2$ in the collection of Z and atoll sources, if confirmed by future observations regarding the measurement of masses, radii, and magnetic field strengths of the neutron stars in LMXBs, can shed light on the nature of interaction between neutron star and accretion flow onto its surface. The observational constraints on the values of $M$, $R$, and $B$ together with the new QPO data, which might be available thanks to the forthcoming missions, could be used to test the model function as far as the distribution of kHz QPO frequencies is concerned. As we have shown in Section~\ref{mffit}, the model function depends on the mass and radius of the neutron star, which in turn define a certain region in the frequency versus $\dot{M}/B^2$ plane where kHz QPOs can be generated only for a specific value of magnetic field strength. If the correlation holds in the current ensemble as suggested by the analyses in Sections~\ref{plfit} and \ref{mffit}, the interaction is then magnetospheric in origin with the Alfv\'{e}n radius, $r_{\mathrm{A}}$, being only one of the basic length scales, which determine the QPO frequencies. The usual assumption that kHz QPO frequencies correspond to the Keplerian frequency or the square of it at the Alfv\'{e}n radius \citep{Zhang2004}, cannot account, however, for the \emph{parallel tracks} in the frequency distribution of an individual source. Moreover, neither $\nu_{1}\propto \nu_{\mathrm{A}}$ nor $\nu_{1}\propto \nu_{\mathrm{A}}^2$ can explain the power-law fit as a fair representative of the average run of lower kHz QPO frequency in the ensemble of sources (Figure~\ref{fig2}c). The presence of another length scale, such as the width of the boundary region where the neutron-star magnetosphere interacts with the accretion disk, can provide us with a possible explanation for the observed distribution of frequencies in an individual source as well as the ensemble of sources (Figures~\ref{fig3} and \ref{fig4}).

The model parameter, $\delta$, represents the radial width of the boundary region in units of $r_{\mathrm{A}}$. As can be seen in Figure~\ref{fig3}, the model function fit to individual source data for the given mass and radius of the neutron star could be obtained if the change in $\delta$ is taken into account. We expect $\delta$ to vary as a function of $\dot{M}$ and therefore $\dot{M}/B^2$ in a source for which the magnetic field strength, $B$, cannot change. The region of the model function fit to data depends on the mass and radius of the neutron star as well as $\delta$. Once $M$ and $R$ are chosen for the source, a unique value for a certain range of $\delta$ can be assigned to the magnetic field strength on the neutron-star surface (Figure~\ref{fig3}). As we have also addressed, in the present work, how $\delta$ is supposed to change as $\dot{M}$ varies in time, we expect $\delta$ to increase with $\dot{M}$ (Figure~\ref{fig5}); otherwise, lower kHz QPO frequencies would monotonically increase to attain extremely high values in the range of 1500--2000 Hz (see Figure~\ref{fig3}) for sufficiently small values of $\delta$ if $\delta$ is kept constant. It is not so difficult to understand why $\delta$ has to increase with $\dot{M}$ at least for a wide range of $\dot{M}$. The radial width of the boundary region at the magnetopause, as we will discuss in a subsequent paper, is determined by the typical half-thickness of the inner disk, which increases with $\dot{M}$ as foreseen by the standard disk model \citep{SS73}. A relatively more elaborate description of the so-called \emph{parallel tracks} in terms of variations in $\delta$ and other physical parameters such as $\Delta \nu$ can be realized within the context of kHz QPO frequencies versus X-ray flux in individual sources \citep{Mendez2000}, provided that the variation of $\delta$ with $\dot{M}$ could be simulated. In a forthcoming paper, we will address the reproduction of kHz QPO data, i.e., the \emph{parallel tracks} of individual sources through modelling the change of $\delta$ with $\dot{M}$.

The additive effect of individual frequency distributions (Section~\ref{mffit}) on the possible correlation between $\nu_{1}$ and $\dot{M}/B^2$ in the ensemble of sources reveals itself as if the frequency data are scattered all along a simple power-law (Figure~\ref{fig2}c) with the power-law coefficient and index being $A\approx 1$ and $\alpha \approx 0.1$, respectively (Section~\ref{plfit}). Our analysis in Section~\ref{mffit} shows that the source distribution in the $\nu_{1}$ versus $\dot{M}/B^2$ plane also depends on the neutron-star masses and radii (Figure~\ref{fig4}). The remarkable similarity between the source distribution in Figure~\ref{fig2}c and the one in Figure~\ref{fig4}b could be an indication of similar neutron-star masses and radii ($M\gtrsim 1M_{\odot }$ and $R\simeq 11$~km) for all sources in the ensemble. It is very likely, on the other hand, that the magnetic field strength varies from one source to another (Table~\ref{table3}). Regardless of the source classification as Z or atoll, we expect to find the frequency tracks with highest slopes, in general, close to the leftmost boundary of the distribution, which is determined by the relatively small value of $\delta$ (see, e.g., Figures~\ref{fig3} and \ref{fig4}). Although there is no strict range for $\delta$, the typical values for its upper and lower limits estimated by the model function fit to data are compatible with those suggested by earlier studies on magnetically threaded boundary regions \citep{EA2004}. The transient source XTE~J1701--462 is unique among all sources in the ensemble as it helps us identify the largest possible range for $\delta$. We could perform the model function fit to the data of other sources using different values of $\delta$, which all lie within this range (Figure~\ref{fig4}).

The exclusion of the single data point of a lower kHz QPO seen at $\sim 500$~Hz during the Z phase of XTE~J1701--462 from the analysis in Section~\ref{nu1cor} is due to the lack of detailed information about the variation of kHz QPO frequencies throughout the Z phase with the source intensity \citep{Sanna2010}. Although the luminosity difference between Z and atoll phases can be deduced from the variation of maximum rms amplitude or quality factor with the luminosity of the source, details of variation of kHz QPO frequencies with luminosity are absent in the Z phase unlike the atoll phase of the same source \citep{Sanna2010}. Despite this uncertainty, the lower kHz QPOs in the Z phase are observed within an extremely narrow range of frequency ($\sim 600-650$~Hz) apart from the QPO at $\sim 500$~Hz. In Section~\ref{nu1cor}, we include all data in the Z phase except this $\sim 500$~Hz QPO. For simplicity, we assume a common luminosity for each data point in this narrow range. The inclusion of the QPO at $\sim 500$~Hz with the same luminosity would be the extreme case with infinite slope for the Z track from $\sim 500$~Hz to $\sim 650$~Hz. Even in this extreme case, it would be possible to describe both Z and atoll phases of XTE~J1701--462 within the region of model function fit to source data. Choosing a smaller value for $\delta$, such as $0.01$ instead of $0.03$ while shifting the data of the source towards lower values of $\dot{M}/B^2$, we would be able to end up with a similar distribution as in Figure~\ref{fig4}. This would only imply slightly higher magnetic field strength for the same mass and radius of the neutron star. In a more realistic case, we would expect to find the QPO at $\sim 500$~Hz with a lower luminosity as compared to others in the $\sim 600-650$~Hz range. The slope of the Z track would then be finite. The present region of the model function fit to data would remain almost the same and yet comprise both Z and atoll phases as in Figure~\ref{fig4}. In any case, the inclusion of the single data point of the QPO at $\sim 500$~Hz wouldn't alter at all the basic outcome of our present analysis.

The existence of the peculiar LMXB XTE~J1701--462 reveals the fact that both Z and atoll characteristics can coexist in the same source provided that the range of luminosity variation is sufficiently large to include values of luminosity as high (low) as those in Z (atoll) sources. The manifestation of both classes in the same source rules out the possibility that the intrinsic source properties such as the mass, radius, spin and magnetic field of the neutron star can play a major role in determining the spectral evolution of the source. The early idea by \citet{HK1989} that the difference between the two spectral classes can be due to the absence (presence) of a magnetosphere in atoll (Z) sources cannot account for the present case of XTE~J1701--462. The transition between high- and low-luminosity phases in the same source suggests that the large variation in $\dot{M}$ can be responsible for the differences between Z and atoll characteristics. The low frequency QPOs, which were detected in Z sources and hence interpreted as the evidence of a magnetosphere, have also been observed in atoll sources \citep{Altamirano2005}.

According to the early arguments in the literature, the magnetic field strength, $B$, on the surface of the neutron star has usually been considered to be the main agent that can account for the differences between Z and atoll classes. In this paper, we claim that the presence of a magnetosphere is not unique to Z phase. In accordance with our present findings regarding the frequency distribution of kHz QPOs, the magnetosphere-disk interaction is also efficient in atoll phase. The kHz QPOs can therefore be generated within the magnetic boundary region, which is presumably stable with respect to specific accretion phases such as Z and atoll. These phases are mainly determined by $\dot{M}$. As shown in Table~\ref{table3}, the surface magnetic field strength of XTE~J1701--462 is less than or close to the value of $B$ in atoll sources such as 4U~1820--30, 4U~1735--44, 4U~1728--34, 4U~1705--44, and 4U~1702--42. This is a very clear evidence of the fact that $B$ has nothing to do with the X-ray spectral state of the source. In addition to XTE~J1701--462, all sources mentioned above are candidates for exhibiting both Z and atoll characteristics if their range of variation in $\dot{M}$ becomes large enough to accommodate the properties of both classes.

The change in the properties of the accretion flow around the neutron star can be the reason for the difference in QPO coherence and rms amplitude between Z and atoll phases \citep{Mendez2006,Sanna2010}. Our analysis suggests that both $\dot{M}$ and $B$ are important in determining the range of kHz QPO frequencies. Unlike the previous arguments in the same context \citep{WZ1997,Zhang2004}, however, we claim that the frequency distribution in an individual source, in addition to the Alfv\'{e}n radius, strongly depends on the length scale of the boundary region, $\delta$, which is also expected to vary with $\dot{M}$. It is highly likely within the model we propose here that $\dot{M}$ has a two-sided effect on QPOs. On the one hand, $\dot{M}$ modifies the length scales, which are mainly responsible for the range of QPO frequencies, on the other hand, it leads to a number of changes in the physical conditions of the accretion flow through the magnetosphere threading the boundary region. The density and temperature of the matter in the magnetosphere can be different in accretion regimes that differ from each other in $\dot{M}$. In this sense, the Z and atoll phases may reflect different accretion regimes. In our picture, the modulation mechanism takes place in the boundary region, which forms the base of the funnel flow extending from the innermost disk radius to the surface of the neutron star. As compared to the soft X-ray emission from the inner disk, the relatively hard X-ray emission associated with the funnel flow may then appear as the frequency resolved energy spectrum characterizing the X-ray variability in LMXBs \citep{Gilfanov2003}.

The possible correlation of lower kHz QPO frequencies with $\dot{M}/B^2$ (Section~\ref{nu1cor}) indicates relatively strong magnetic fields for Z sources in comparison with atoll sources (Table~\ref{table3}). Apart from XTE~J1701--462, which exhibits both Z and atoll characteristics, this tendency may have a simple explanation (being different from the arguments based on magnetic field for a direct explanation of Z and atoll characteristics before the emergence of XTE~J1701--462). The fact that some sources, which are observed only in Z phase (high $\dot{M}$) and never in atoll phase, such as Cyg~X-2, GX~5--1, GX~17+2, and Sco~X-1, could be due to the propeller effect that would arise when the Alfv\'{e}n radius exceeds the co-rotation radius at sufficiently low mass accretion rates. If so, then we would expect to observe most of high $B$ sources in high $\dot{M}$ regime in accordance with the present statistics regarding the number of Z sources in comparison with that of atoll sources within the population of neutron star LMXBs.

\acknowledgments We thank M.~A. Alpar and D. Psaltis for reading the manuscript and useful suggestions. We also thank the anonymous referee for important issues, useful comments and suggestions that led us to improve this manuscript. This work was supported by the Scientific and Technological Research Council of Turkey (T\"{U}B\.{I}TAK), under the project grant 114F100.

\end{document}